\documentclass[12pt,article]{aastex}

\usepackage{amssymb,graphicx,latexsym,multirow,rotating,url}
\newcommand{\fpos}{\ifmmode{f_{\!\mbox{\footnotesize \it pos}}}\else{$f_{\!\mbox{\footnotesize\it pos}}$}\fi}

\newcommand{\Ha}{{H$\alpha$}}
\newcommand{\kmps}{km~s\ensuremath{^{-1}}}
\newcommand{\eg}{{\it e.g.,}\/}

\newcommand{\ie}{{\it i.e.,}\ }

\begin{document}
\noindent
\textbf{\Large \\ Chapter 4: \\ Derivations and Observations of Prominence Bulk \\ Motions and Mass}\footnote{In  \textit{Solar Prominences}, eds. J.-C. Vial \& O. Engvold, Springer (2015)}

\noindent
T.A.\ Kucera

\noindent
NASA Goddard Space Flight Center, Code 671, Greenbelt, MD 20771, USA

\section*{Abstract}
In this chapter we review observations and techniques for measuring both bulk flows in prominences 
and prominence mass. Measuring these quantities is essential to development and testing of 
models discussed throughout this book. Prominence flows are complex and various, ranging from 
the relatively linear flows along prominence spines to the complex, turbulent patterns exhibited by 
hedgerow prominences. Techniques for measuring flows include  time slice and optical flow 
techniques used for motions in the plane of the sky and the use of spectral line profiles to determine 
Doppler velocities along the line of sight. Prominence mass measurement is chiefly done via 
continuum absorption measurements, but mass has also been estimated using cloud modeling and white 
light measurements.

\section{Bulk Motions}
\label{sec:5}

Prominences have long been known to be dynamic structures, displaying internal
motions of various kinds even when globally at rest. Here we discuss  bulk motions 
and methods used for measuring them, although the measurement methods are also 
relevant to the oscillatory motions discussed in Chapter 11~\citep{ballester_15}.

A good understanding of flows is highly important for testing and constraining
models of prominence formation and stability, determining the role of flows in
the force and energy equilibria of prominences, and casting light on closely
related questions concerning prominence magnetic field structure.  
For instance, evaporative-condensation models predict that flows of 
cool material should originate in the corona, while injection and levitation 
models involve cool material flowing upwards form the chromosphere. 
If we assume that flows are moving along magnetic field lines then the flow trajectories can 
tell us about the magnetic structure of the prominence, but it could also 
be the case that the material is moving with a changing field or even in 
some cases diffusing across the field. The various theoretical models of 
prominence magnetic field and related dynamics and their predictions are 
discussed in more detail in Chapter 10~\citep{karpen_15}. 

To obtain the required information we need detailed trajectories of prominence
plasma features, including the origin of the plasma, and any changes in
temperature and velocity.  How do observed motions compare to those expected 
from various models of the prominence magnetic field and actual magnetic 
field measurements?  What connections,  if any, are there between 
flows at different temperatures? What processes 
can explain these observed flows?

In the last few decades, new instrumentation has yielded more information
concerning flows. In the visible regime, the new high temporal and spatial
resolution data combined with spectral information have revealed  
counter-streaming flows in filaments on the disk in observations such 
as those from the the Swedish Solar Telescope (SST) at La Palma. The similarly good spatial resolution
of the \textit{Hinode}/Solar Optical Telescope (SOT) combined with excellent 
long term image stability allowed by its space based platform
have allowed detailed studies of the complex motions observed in prominences 
at the limb.

In the UV and EUV range spectrographs  such as \textit{Solar and Heliospheric Observatory (SOHO)'s}  
Coronal Diagnostic Spectrometer (CDS) and Solar Ultraviolet Measurements of Emitted Radiation (SUMER) 
and \textit{Hinode's} EUV Imaging Spectrometer (EIS) have provided information in a range of lines 
formed at chromospheric, transition region, and coronal temperatures. The new \textit{Interface Region 
Imaging Spectrograph (IRIS)} mission provides higher resolution spectroscopic 
as well as plane-of-sky information concerning the motions of plasma at chromospheric and
transition region temperatures. High cadence UV and EUV imaging information
from the \textit{Solar Dynamics Observatory (SDO)'s} Atmospheric Imaging Assembly (AIA) and, 
earlier, from the \textit{Transition Region and Coronal Explorer (TRACE)}  and \textit{SOHO}/Extreme 
ultraviolet Imaging Telescope (EIT) have been important as well.

\subsection{Measurements of Motions}

Motions in the plane of the sky are measured by tracking actual features, while
line of sight (LOS) motions are detected using Doppler shifts.  Sometimes the 
plane-of-sky (POS) and line position methods can be combined to good effect. 
With either LOS or POS measurements alone, we have no 
direct information on the 3D structure. However, in some cases we can make
estimates based on the knowledge of the orientation of prominence features as
viewed over several days or from two points of view using the \textit{Solar Terrestrial 
Relations Observatory (STEREO)} spacecraft.

\subsubsection{Motions in the Plane of the Sky}
Observations of motions in the plane of the sky have the advantage that it is
possible to pick out actual moving features, making it straight forward in some 
cases to track the motion by simply marking off the feature location in successive images. 

There are a number of general things to be careful of, however. The very fact that such 
measurements are feature driven makes it possible to miss evenly moving flows that vary 
below the resolution of the data.
Also, there is the question of what is actually moving. Often a variation in intensity may indeed be 
due to actual moving material, but it is also possible it is due to a propagating disturbance in 
density or temperature. There are cases, especially in optically thin plasmas, in which multiple 
layers of plasma can produce confusing intensity variations and make the data hard to 
interpret. It is important to have sufficiently high spatial and temporal resolution to avoid 
aliasing and be able to reliably identify and track particular features.  
One should also consider projection effects. For
instance, when a prominence is seen along the limb it can be difficult to
distinguish vertical motions from horizontal flows curving over the solar limb.

\begin{figure}
\includegraphics[angle=90,width=0.5\textwidth, trim=0 140 0 80]{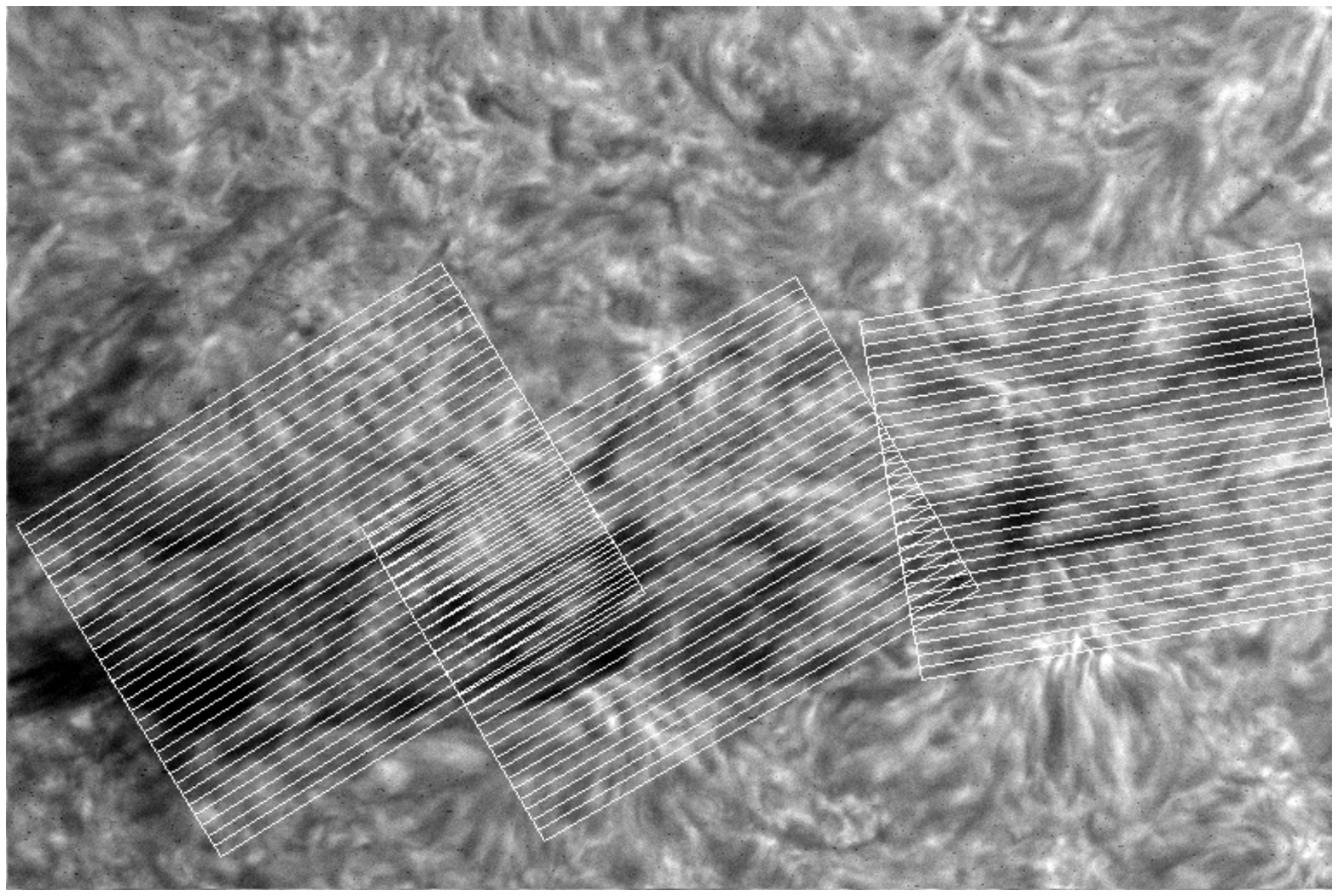}
\includegraphics[width=0.45\textwidth]{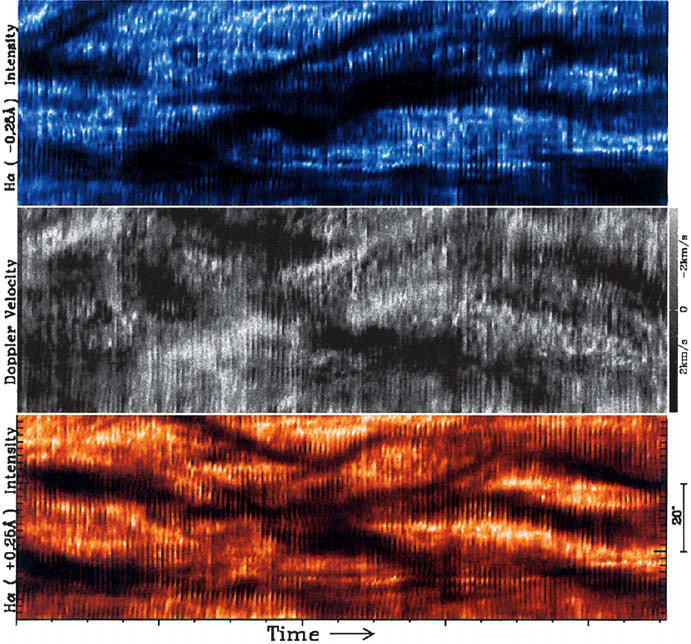}
\caption{An illustration of the use of time slice diagrams to analyze flows in
a \Ha\ filament observed  with the SST~\citep{lin_03}. The left panel shows the  locations of
parallel slices used for analyzing velocities in \Ha\ fine structures. At right
from top to bottom are the blue wing intensities, Doppler shift, and red wing
intensities along a single time slice located in the middle section of the filament. Diagonal 
features show motions along the slice. 
 \label{f:Lin03}}
\end{figure}

\textbf{Time-Slice}
A common variation of feature tracking is to use the time slice method (see
Fig.~\ref{f:Lin03}). The intensity or velocity along a trajectory traced out in consecutive images of
the filament is followed versus time, and the slope of the brightening or
darkening in this 2D diagram gives a measurement of the velocity of the feature
in the plane of the sky. This method can be easily applied to high spatial
resolution images, \eg\ with the SST or the Dutch Open
Telescope (DOT) at La Palma, and has also been used with EUV images. The slices
can be linear, but they can also curve to follow the trajectory of a particular
feature.

\textbf{Optical Flow Techniques}
There has also been some use of automatic tracking to trace prominence flow
fields. This can be difficult because of the complex nature of prominences,
often including multiple sources that may be optically thin or combine both
emission and absorption. The techniques that have been used successfully are
best applied to areas that can be described in terms of flow fields rather than
discrete moving features.

One of the most commonly used of these methods is Local Correlation Tracking
(LCT)~\citep{leese_70}, in which a two dimensional cross-correlation is applied to
a pair of images to determine the shifts for different parts of the images so as to
maximize the function~\citep{november_88}
\begin{equation}
C(\delta,x)=\int J_t(\xi-{\delta\over2})J_{t+\tau}(\xi+{\delta\over 2})W(x-\xi) \partial\xi
\end{equation}
where $J_t(x)$ and $J_{t+\tau}(x)$ are the two images taken at times $t$ and
$t+\tau$ respectively, with a vector displacement $\delta$ between them. $W(x)$
is a windowing function that controls the size of the region over which the
images are compared. 

\begin{figure}
	\center
\includegraphics[width=.7\textwidth]{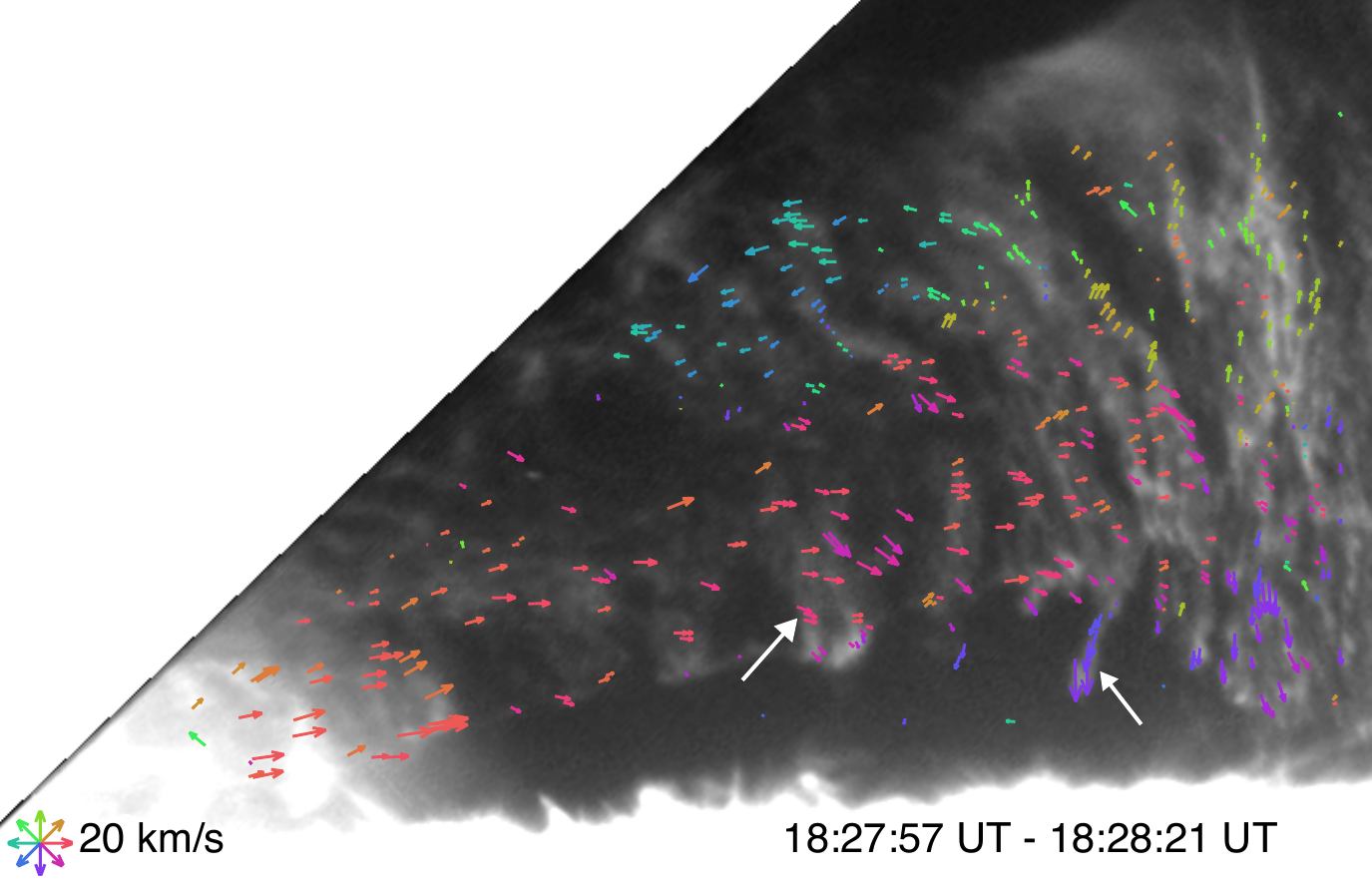}
\caption{Vertically oriented features seen in a hedgerow prominence  observed with 
\textit{Hinode}/SOT {in the \Ha\ line}~\citep{chae_08a}. 
They appear to move both horizontally and vertically in the plane of the sky. White 
arrows point out a newly formed feature and one which is falling downwards.}
\label{f:Chae08}
\end{figure}

Also used are optical flow techniques originally developed
for magnetogram analysis, such as the nonlinear affine velocity estimator
(NAVE)~\citep{schuck_06} which was used to track the flows shown in
Fig.~\ref{f:Chae08}.  These algorithms utilize spatial derivatives of the images
and allow for more complex local flows~\citep{chae_08b}.

\subsubsection{Along the Line of Sight: Doppler Observations}
Doppler measurements more reliably give access to steady flows.  If there is
sufficient spectral resolution the Doppler velocity is generally determined by
fitting the observed spectral line, usually with a Gaussian function, and
applying the standard formula for Doppler shifts,
\begin{equation}
v/c=(\lambda-\lambda_0)/\lambda_0
\end{equation}
where $\lambda_0$ is the rest wavelength of the line, $\lambda$ the line wavelength of the source, $v$ the
source velocity, and $c$ the speed of light. 

When specifying a Doppler velocity it is important to be clear how the zero velocity point is 
determined.  In many
cases one cannot assume that the data one is using has been calibrated with
respect to the absolute wavelength. The wavelength scale often shifts in the
instrument due to instrumental heating or, for ground based instruments, 
changes in the refractive index of the atmosphere. 
Although it is possible for some wavebands to incorporate a wavelength standard into an instrument this
is not often done in space-based solar spectrographs.

One method of velocity calibration is to select a feature in the field of view or that has been
recently observed and declare it to be at rest so that the velocity values are
relative to that feature. For spectral observations in visible wavelengths one
should optimally precede or follow one's observations with observations of a
quiet area at disk center. The average spectral profile over the region is then
used to determine the rest wavelength position. The wavelength scale is also
sometimes calibrated by measuring it compared to certain spectral lines that
are assumed to be at rest. For instance, for UV or EUV observations of the solar disk absolute
wavelength calibrations are often done by assuming that chromospheric spectral
lines averaged over a large area are at rest.  If the Doppler shifts of interest 
are to be measured using a
spectral line produced at chromospheric or transition region temperatures observed above the limb 
(as is the case for a prominence), a coronal line at a near-by wavelength (again averaged over a large area) 
might be used as a standard. However, coronal
plasma may not be at rest either. Thus one should be aware of the uncertainties
in the chosen method and consider how well the absolute Doppler shift
has been measured.

\begin{figure}
\includegraphics[width=.5\textwidth]{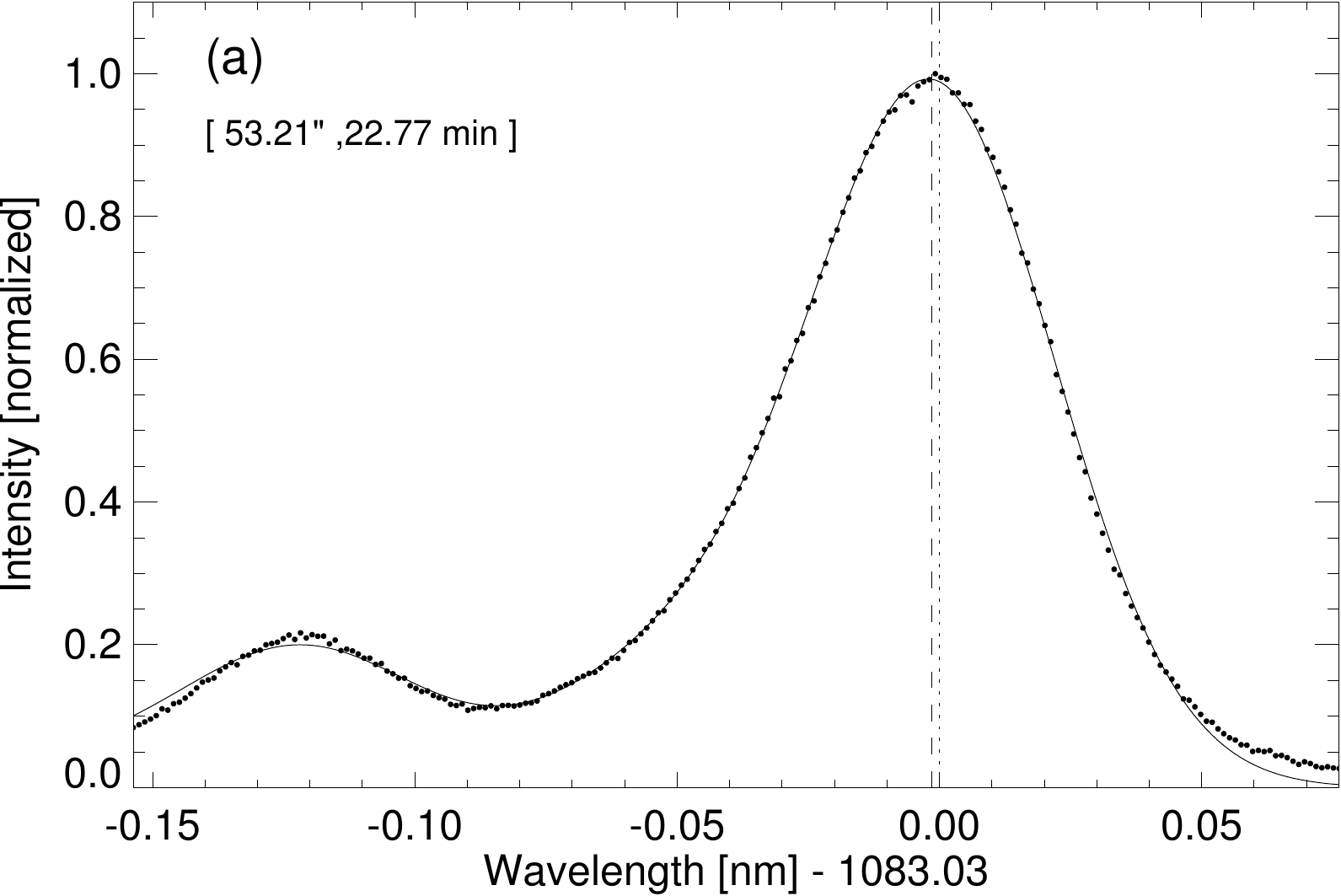}
\includegraphics[width=.5\textwidth]{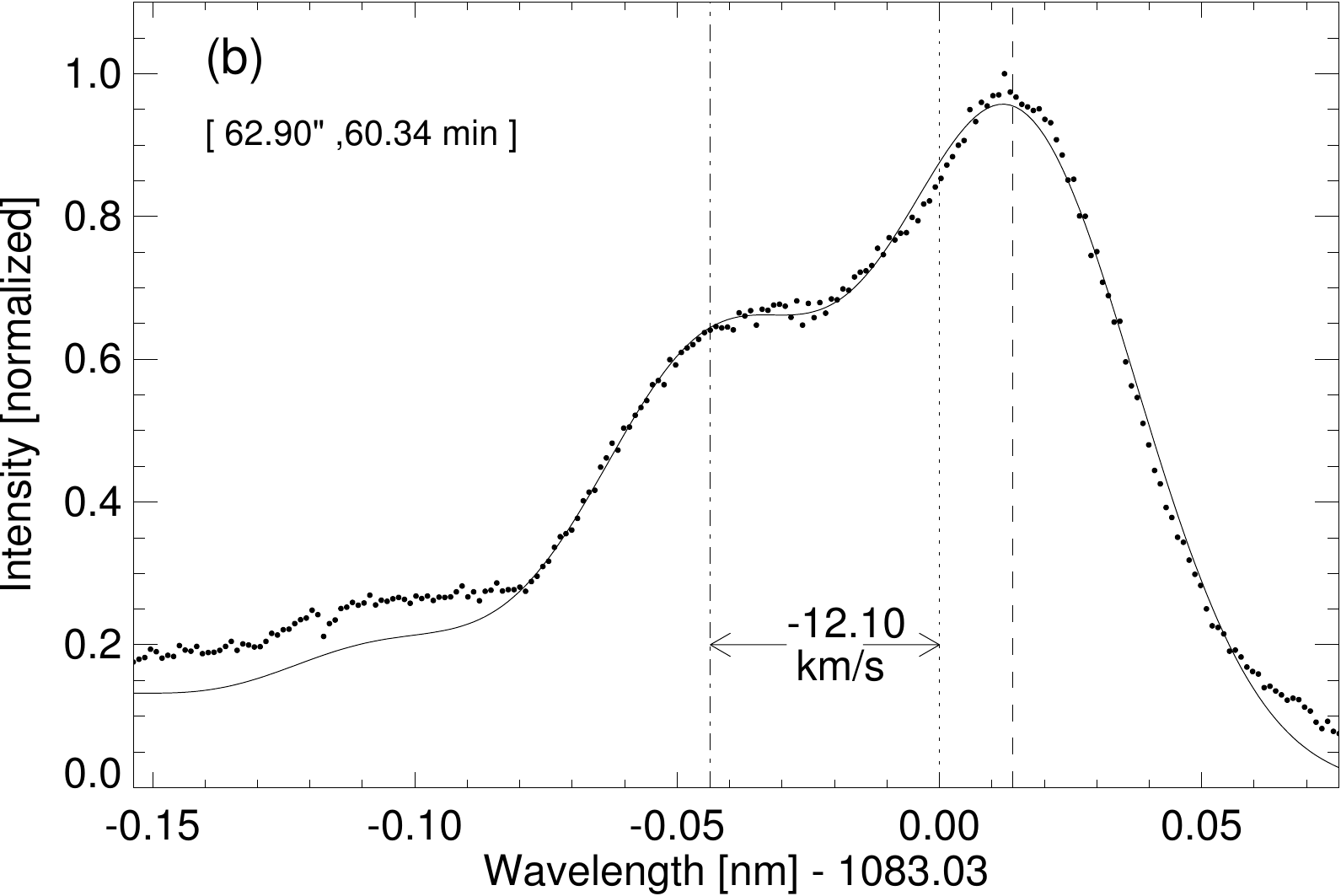}
\caption{Example of a multi-component Doppler shift in a prominence footpoint in 
He I 10830~\AA~\citep{orozco_12}. (a) shows a rest profile consisting of a weak
line at 10829.09~\AA\ (due to the $^3S_1$-$^3P_0$ transition) and the main line at 
10830.29~\AA\ ($^2S_1$-$^3P_1$ and $^2S_1$-$^3P_2$). (b) shows a spectrum with two Doppler components in the main line, one red shifted and one 
blue shifted. A spectral fit with a single component might 
indicate little or no motion along the line of sight. The measurements were made with data from Tenerife Infrared Polarimeter attached to the German Vacuum Tower Telescope.}
\label{f:TwoDopComp}
\end{figure}

A spectrally resolved profile is generally fit with some function, usually a Gaussian. 
Often there is more than one Doppler component present. This is
especially likely in optically thin plasmas, where the emission is an integration of the LOS
velocity of many different features so that the measurement provides a 
value averaged along the LOS. It is possible, for instance, for some of the plasma in the
field of view to be relatively stationary and for other parts of it to be
moving. An example is shown in Fig.~\ref{f:TwoDopComp}. Here, fitting the entire 
spectral line as a unit could result in a measurement of almost no shift at all. 
At best one would be left with the impression of a broadened line profile with 
no information concerning the details of the flow. Instead, the line has been fit by 
two Gaussian functions, indicating a blue shifted component moving at 
12~km~s$^{-1}$ and a red shifted component at about 4~km~s$^{-1}$~\citep{orozco_12}.

A related issue is that of line blending, in which multiple lines are
present. If all the lines are from the same source they will move in tandem, but
if they are not (if, for instance, they are formed at different temperatures in
different regions) fitting the line as a single source will again yield
incorrect results. Even if two blended lines are formed from the same ion, a 
density dependence in their ratio can lead to spurious Doppler signals.

Another aspect to Doppler line fitting is that of the reversed lines, which is
particularly relevant to prominences. Some optically thick lines exhibit central
reversals. Full modeling of the behavior of such lines is discussed in Chapter
5~\citep{heinzel_15}, but the location of the center of the absorption feature can be an indicator of
the relative velocities of different parts of the structure.

For observations of filaments on the disk,  the signal coming from the
prominence must be disentangled from the chromospheric  background. Various
techniques have been developed to do this, principally based on cloud model
methods~\citep{beckers_64,mein_96,tziotziou_07}.  With such methods, the data is fit to a  
non-LTE model model of the prominence as a ``cloud'' suspended above the 
chromosphere utilizing four parameters:  the LOS velocity $V$,  the source function $S$, the
optical thickness $\tau$, and  the line width $\Delta \lambda _{D}$.  $S$ and
$\tau$ are strongly coupled but $V$ can be  computed. Cloud models are described 
in more detail in Chapter 5~\citep{heinzel_15}

For all these caveats and difficulties, there are short-cuts that are taken. 
Lines are sometimes sampled at only two or three
wavelengths. Differences in the intensities of the red and blue wings of the lines
are then used to identify general regions of flow or to separate out material 
flowing in different directions.

Optimally measurement of flows in and transverse to the plane of
the sky can be combined to reveal the three dimensional trajectory of the
motion. One particular method used to combine the two is to trace motions using an
image slice method applied to the red or blue wing of a spectral line.   An
example of this is shown in Fig.~\ref{f:Lin03}, in which a time slice analysis
is done in the red and blue wings of the \Ha\ line, which show different
features~\citep{lin_03}.

\subsection{Observations of Flows in Prominences}
Prominences show a variety of different flows. Here we describe some 
of the characteristic ones for non-erupting prominences. Eruptions and 
eruption precursors are discussed in Chapters 15 and 16~\citep{webb_15,gopalswamy_15}.

\subsubsection{Flows in Quiescent Prominences}
\label{s:QPromFlows}
What are referred to as quiescent prominences exhibit a range of flow behaviors
including motions of vertically aligned structures in hedgerow prominences and
also the motions in the spine and barb formations of lower latitude prominences
in quiet regions. Although traditionally quiescent prominences are thought of as
relatively stationary, even the most stable quiescent filaments are dynamic structures
exhibiting flows and oscillations.

\begin{figure}
 \begin{minipage}[c]{0.6\textwidth}
\includegraphics[width=\textwidth]{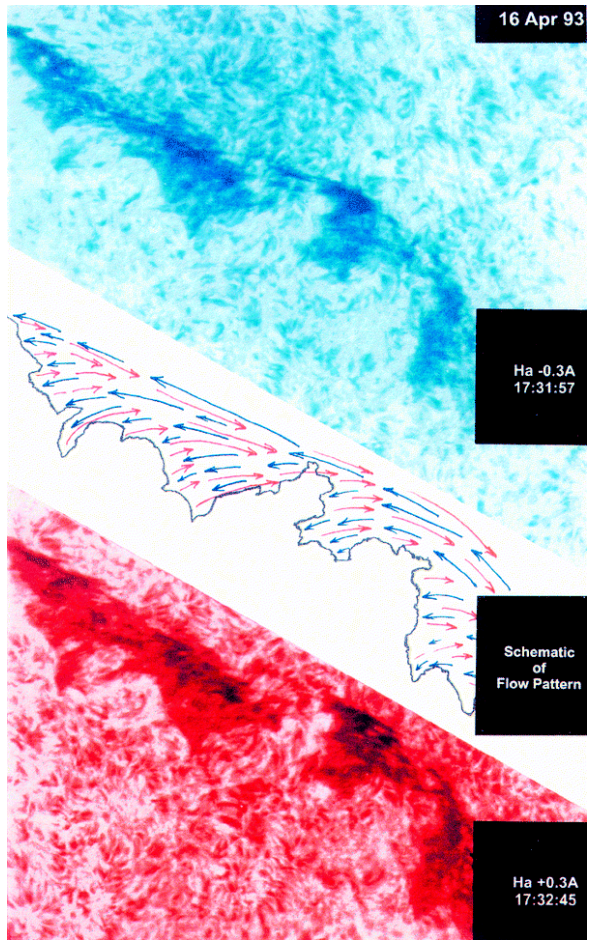}
\end{minipage}
\begin{minipage}[c]{0.4\textwidth}
\caption{Counter-streaming observed in along a prominence spine and barbs observed at 
Big Bear Solar Observatory~\citep{zirker_98}. The flows were detected by constructing movies 
from images in the blue wing (top) and red wing (bottom) of the \Ha\ line.}
\label{f:zirker98}
\end{minipage}
\end{figure}

\textbf{Spine flows}
Quiescent prominences not in the polar crown often show a two-part structure of
a long spine accompanied by barbs which extend down to the chromosphere~\citep{engvold_15}.  
Observations in \Ha\ show flows 
along prominence spines and barbs at speeds  of 10-20~\kmps. As shown in the 
schematic in Fig.~\ref {f:zirker98}, the flows can go in both directions 
simultaneously along the spine, a phenomenon known as
counter-streaming~\citep{zirker_98}. These observations report individual moving
features traced over distances of 10,000 to 100,000~km. 
Other observations of spine flows in an intermediate prominence show moving features changing direction, suggesting counter-streaming may be the result of plasma oscillating along the magnetic field~\citep{ahn_10}.

Although polar crown prominences often appear to consist mostly of barbs when
observed in \Ha,  observations in the 304~\AA\ EUV imaging band show flowing material
along spines, evidently too hot or optically thin to be observed in \Ha\, but apparent
in the more optically thick He~II 304~\AA\ line.  

He II 304~\AA\ emitting plasma
has been observed moving along prominence spines at speeds up to 75~km~s$^{-1}$ in the plane of the
sky~\citep{wangym_99}, much faster than considered typical for \Ha\ flows,
and similar observations have been made in other transition region temperature
lines in the EUV~\citep{kucera_03}.

\textbf{Flows in Barbs}
Although barbs in polar crown prominences often take the form of hedgerows (see below), 
in lower latitude prominences they are narrower structures, appearing as short 
outgrowths from the spine against the disk in \Ha\ and as thin dark 
pillars in EUV images of prominences along the limb. These EUV pillars exhibit swaying motions that can be 
interpreted as material oscillating back and forth~\citep{panasenco_14}, perhaps in 
dipped field lines. Alternatively, it has  been suggested based on Doppler 
data that these motions may be due to  helical motions around a roughly 
vertical axis~\citep{orozco_12}. This seems hard to reconcile with the appearance 
of these structures from above. From that point of view they appear to consist of relatively straight threads.

\textbf{Hedgerow Prominence Flows} \label{s:hedgerow}
Hedgerow prominences~\citep{engvold_15} tend to form in the polar crown. As mentioned above, hedgerows as seen in \Ha\ are often interpreted as large barbs, with the prominence spine not strongly visible in \Ha\ but observed in \ion{He}{2} 304~\AA\ line.

Hedgerows barbs as seen in \Ha\ often show structures that are aligned perpendicular to the solar limb.
 These structures have been observed for many years, but recent high resolution
observations have renewed focus on these features and related motions. 

\begin{figure}[b!]
	\center
\includegraphics[width=.6\textwidth]{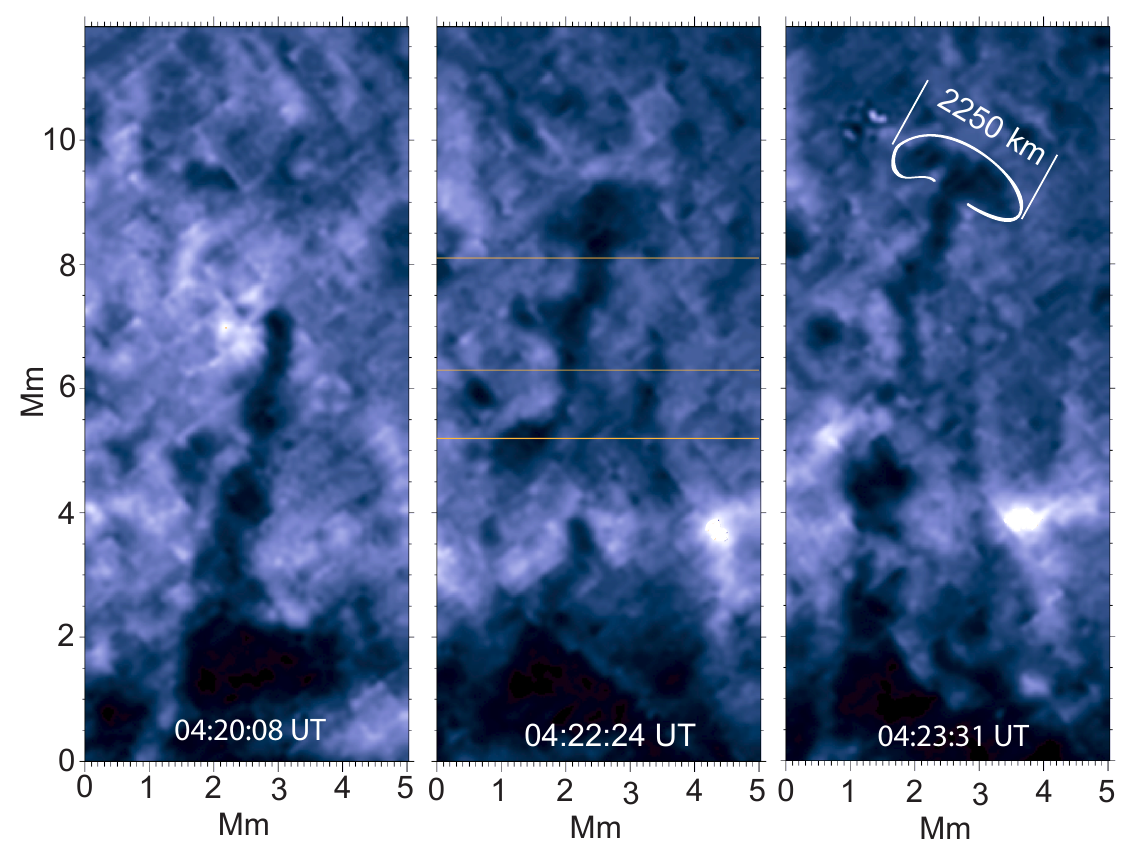}
\caption{Dark upward moving feature with a ``mushroom cap''  head observed on 
the limb with \textit{Hinode}/SOT in the Ca~II H (3968~\AA) line, ascending with a mean 
velocity of 22.9~\kmps\ in the plane of the sky~\citep{berger_08}. }
\label{f:Berger08}
\end{figure}

The \textit{Hinode} SOT instrument in particular has made possible images of prominences
and prominence flows at the limb with unprecedented resolution and stability.
Observations of large, quiescent hedgerow prominences in the \ion{Ca}{2} H and \Ha\  
lines show vertically aligned bright and dark features. An example of the bright 
features is shown in Fig.~\ref{f:Chae08}, which shows a series
of bright vertical structures exhibiting vortical-type motions. They move horizontally at 
about 10~\kmps\ and then move downwards so that individual blobs attain speeds of 35~\kmps.  
The downwards acceleration of the blobs is $0.015$~km~s$^{-2}$ or less, significantly below the gravitational acceleration of 0.27~km s$^{-2}$~\citep{chae_08a}. Also observed are turbulent looking dark features moving upwards from the bottom edge of the hedgerow (Fig.~\ref{f:Berger08}). These plumes have maximum initial speeds in the plane of the sky of 20-30~\kmps\ and decelerate as they rise~\citep{berger_10}.

 A  number of possible explanations for the sub-gravitational acceleration in hedgerows have been suggested.
This phenomenon is also observed in coronal cloud prominences and coronal rain (see \S~\ref{s:rain}), 
although it is not clear if the same phenomena is at work. Possible mechanisms include density 
enhancements in vertical magnetic fields~\citep{mackay_01},  Lorentz forces associated with 
small scale horizontal or tangled fields~\citep{low_05,vanballegooijen_10}, and pressure from 
waves transverse to the field~\citep{pecseli_00,antolin_11}.

Doppler data indicate that the apparently vertical structures in hedgerows also have a 
velocity component along the line of sight~\citep{schmieder_10}. This may be relevant 
to both the sub-gravitational acceleration and the
question of how these features on the limb relate to those seen on the disk. 
Contrary to the rather turbulent appearance of hedgerows observed from the side,
\Ha\ observations from above  show relatively direct flows along straight thread-like 
structures with inclined up- and down-flows in the barbs. It is not clear how these 
two points of view are to be reconciled. 

\subsubsection{Flows in Active Region Prominences}
\label{s:ARPromFlows}

\begin{figure}[b!]
\begin{minipage}{\textwidth}
  \centering
  $\vcenter{\hbox{\includegraphics[width=.4\textwidth]{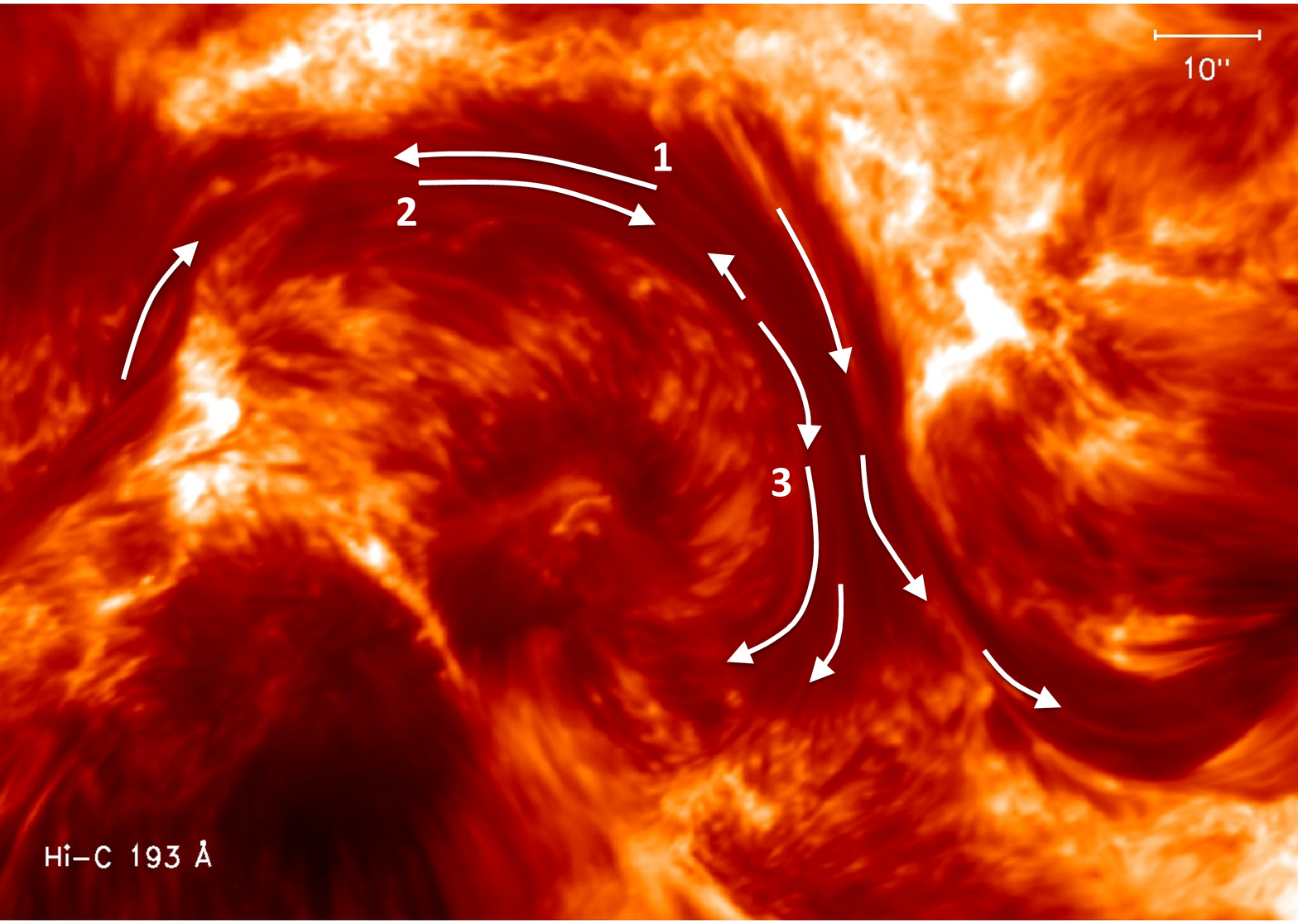}}}$
  \hspace*{.2in}
  $\vcenter{\hbox{\includegraphics[width=.4\textwidth]{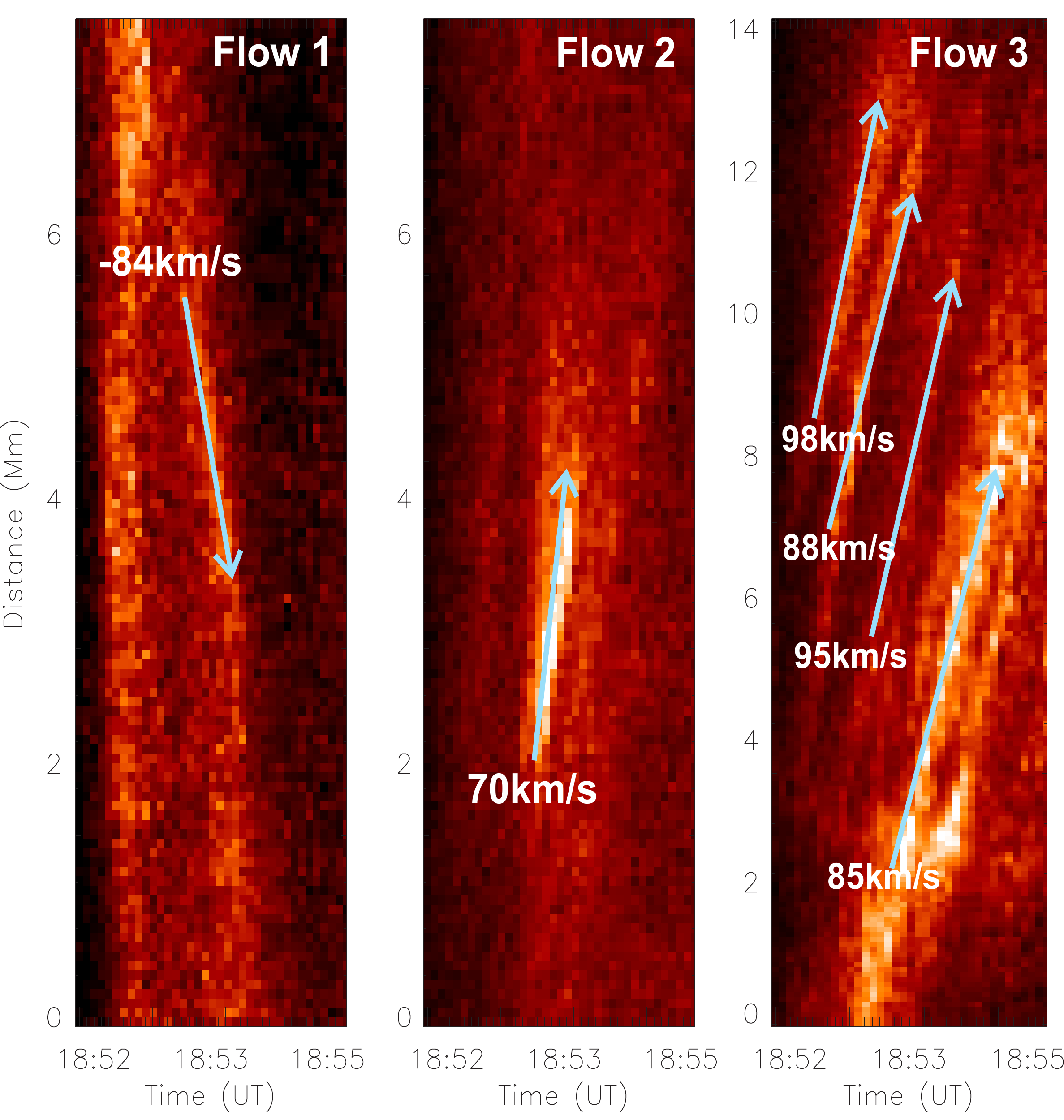}}}$
\end{minipage}
\caption{Flows observed in an active region filament with the rocket-borne 
High-resolution Coronal Imager (Hi-C) instrument in a 193~\AA\ band~\citep{alexander_13}. 
Flows were observed in both directions along the spine with velocities in the plane of 
the sky between 70 and 100~\kmps.}
\label{f:HiC}
\end{figure}

One of the signatures of active region prominences is their high level of
activity, which includes easily seen flows along the prominence axis. A number
of studies report the formation of active region prominences via abrupt jet-like
flows, often associated with observed activity in the magnetic field. Velocities
in forming active region filaments can be as high as 250~\kmps~\citep{chae_03}.
Even once formed the regions show flows along the spine, which include
counter-streaming motions in which there are multiple flows in the spine moving
in opposite directions. Fig.~\ref{f:HiC} shows motions of almost 100~\kmps\ in an 
active region filament observed in the 193~\AA\ band~\citep{alexander_13}.
In addition to these flows observed in cool prominence
plasma, active regions filaments are also associated with coronal temperature
jet-like features directed parallel to or spiraling around the cooler filament
material.

 \subsubsection{Coronal Cloud Prominences and Coronal Rain}
 \label{s:rain}
Coronal cloud prominences are characterized by drainage along curved
trajectories. One study from the 1970's found acceleration in some coronal 
cloud prominence flows to be consistent with that expected from 
gravitation~\citep{engvold_76}. More recent observations report accelerations
significantly less than gravitational acceleration~\citep{stenborg_08,liuw_12}.
As discussed in \S\ref{s:hedgerow}, a number of theories have been 
put forward to explain such sub-gravitational acceleration, 

The draining of chromospheric temperature material from the corona is know 
as ``coronal rain.'' Coronal rain is also commonly observed in unstable loops 
above active regions. Combined EUV and visible observations of such loops 
show that the phenomena is due to hot, coronal temperature loops cooling, 
resulting in the raining down of condensing material~\citep{schrijver_01}.
This may be due to the sort of evaporation-condensation process 
hypothesized for prominences~\citep{karpen_15} 
but applied to arched field lines instead of the horizontal or dipped field lines 
associated with prominence inversion lines. As in coronal cloud prominences,
the observed acceleration is less than that expected from gravity. Average 
speeds are about 70~km~s$^{-1}$~\citep{antolin_12}.

\subsubsection{Flows Observed in the PCTR and Corona}
\label{s:EUVPromFlows}

In many ways, observations of the prominence corona transition region (PCTR) 
and corona in  and around prominences 
are quite limited compared to \Ha\
observations  since they rarely reach the combination of high temporal, spatial,
and sometimes spectral resolutions available in \Ha\ (although IRIS should offer
an improvement on this for some transition region lines). However, the insight
they provide with regard to temperature  makes them important to our
understanding of basic prominence properties.

One key question is how hotter temperature plasma is moving relative to the 
cool material forming the prominence core. Some observations do show emissions formed at
different transition region temperatures from what appear to be the same moving
source, especially below about
$2.5\times10^5$~K~\citep{wiik_93,kucera_03,kucera_06, kucera_08},  suggesting a
cool core with a PCTR accompanying it. 
Other studies, especially Doppler and related modeling, have suggested that
there may be different threads formed at different temperatures with a range of
velocities~\citep{deboer_98,cirigliano_04}.

As described above, EUV observations of transition region temperature material in 
prominences have shown faster moving plasma than what is considered normal in  
\Ha.  This may be because they highlight portions of the prominences not observed 
in \Ha. However, even observations of  absorption near 195~\AA, which 
should reflect the same plasma as \Ha, report quite fast motions~\citep{panasenco_14}. 
This may be a selection effect resulting from the relatively low spatial resolution of most 
EUV imagers as compared to the highest resolution \Ha\ telescopes.

The higher temperature transition region lines and even corona lines highlight the 
extent to which the prominence is part of a larger magnetic structure. In active 
prominences bright jets of coronal temperature material spiral around prominence spines. 
Even quiescent prominences viewed in EUV at the limb along direction of the prominence spine
reveal flows along apparently spiral tracks inside prominence cavity, while the cool dense 
prominence material collects along the bottom. 
Flows in cavities associated with prominences are discussed in detail in Chapter 13~\citep{gibson_15}.

\subsubsection{Flows on Time Scales of Days}
Prominences evolve on different time scales. For instance, they are been 
measured to rise and expand with time, especially in the days before eruption~\citep{liuk_12}. 
Such changes are presumably due to evolution of the magnetic field in which the 
prominence is embedded~\citep{gibson_15}.

Another sort of slow evolution has been reported in stable filaments. 
Kilper et al.~\citep{kilper_09} report that such filaments show a decrease in 
ratio of \ion{He}{1} 10830~\AA\ to \Ha\  emission as a function of 
height over time, although this variation disappears 
in more active or erupting filaments. They interpret this as sign of cross 
field diffusion which would be expected to be much faster for the heavier 
helium atoms than for hydrogen~\citep{gilbert_07}.

\section{Prominence Mass}
\label{s:mass}
Prominence mass is an important quantity in our understanding of prominence 
plasma - a basic physical quantity to be accounted for by models of prominence 
formation and also eruption. Here we will discuss methods used for measuring 
prominence mass and summarize some of the results. 

\subsection{Continuum Absorption}
\label{s:LyCont}
\begin{figure}[b!]
\includegraphics[width=0.9\textwidth]{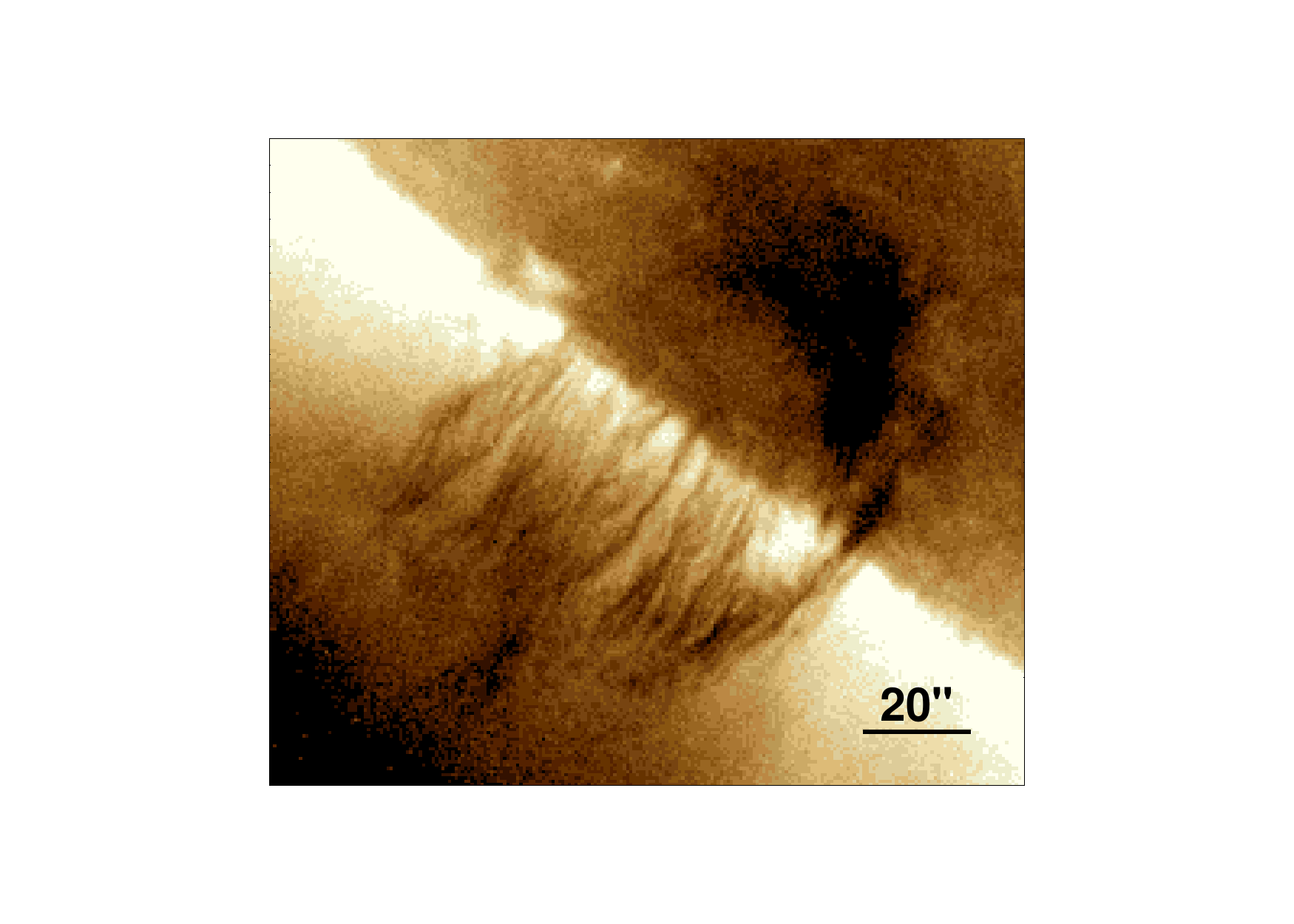}
\caption{Prominence observed in absorption in the 193~\AA\ band of \textit{SDO}/AIA on 2010 August 11 at 20:00~UT.}
\label{f:LymanProm}
\end{figure}

\begin{figure}
\includegraphics[width=8cm]{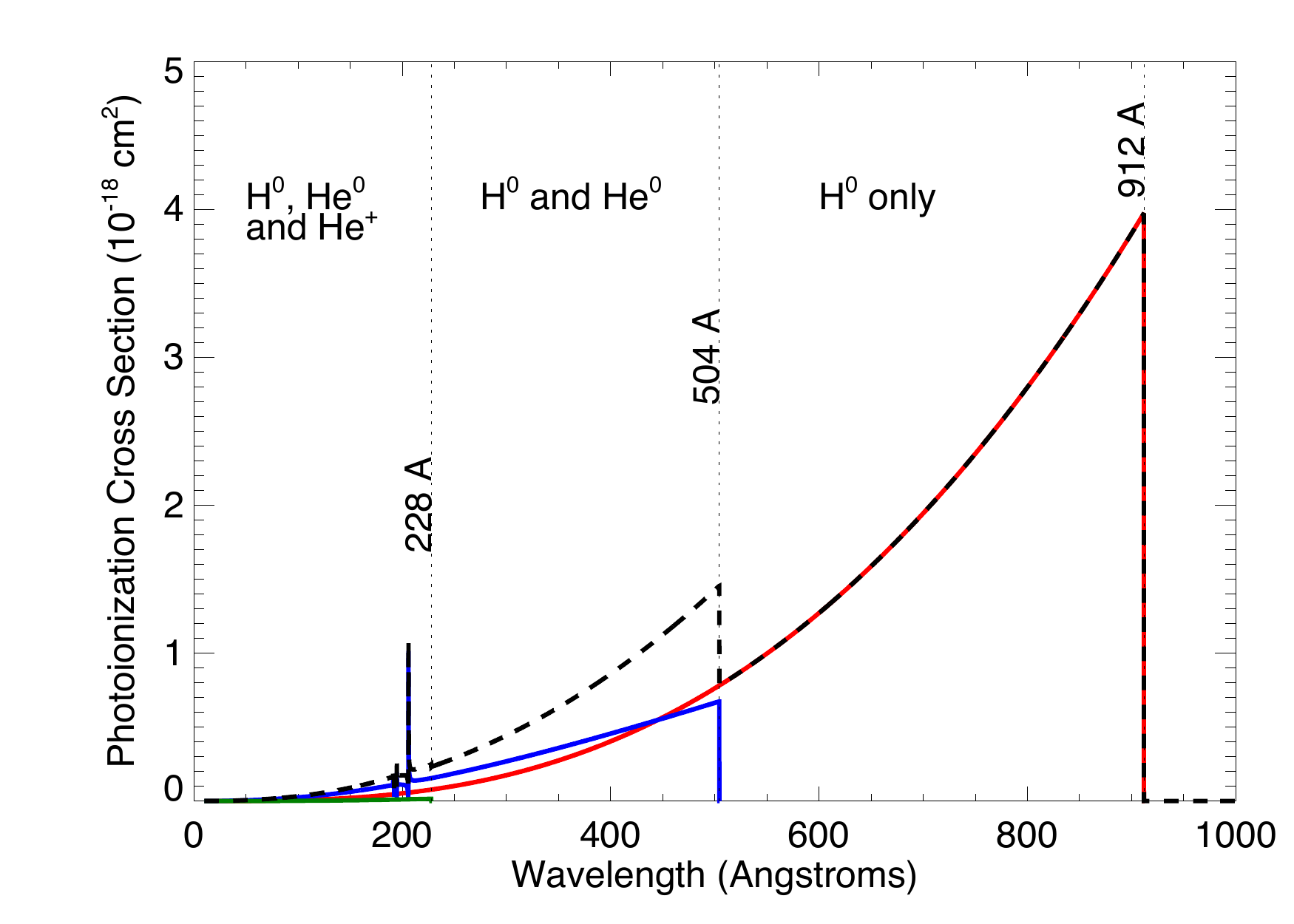}
\caption{Photoionization cross-section per atom/ion in a prominence of assumed hydrogen 
fraction 0.9, hydrogen fractional ionization 0.3 and helium ionization fraction 0.1. Neutral 
hydrogen~\citep{karzas_61} (red) contributes below 912~\AA, He$^0$~\citep{fernley_87} (blue) 
below 504~\AA\ and He$^+$~\citep{karzas_61} (green) below 228~\AA. The He I curve shows 
the locations of the helium auto-ionization features. The total average is shown by the dashed line. 
Code courtesy V.\ Andretta. }
\label{f:LymanXSection}
\end{figure}

The most frequently used  method for measuring prominence mass is via
measurements of continuum absorption.  An example of a prominence seen in
absorption in EUV is shown in Fig.~\ref{f:LymanProm}. This is absorption due to the photoionization of
neutral hydrogen and neutral and once ionized helium, chiefly Lyman-absorption from the
ground state, and is proportional to $\exp{(-\sigma N)}$ where $N$ is the column number 
of absorbing atoms or ions and $\sigma$ is the absorption cross section.
As shown in Fig.~\ref{f:LymanXSection}, the absorption cross section for each of these 
species has an upper limit in wavelength, 912~\AA\
for H$^0$, 504~\AA\ for He$^0$, and 228~\AA\ He$^+$ below which it decreases.
There are also auto-ionization resonances in the neutral helium cross section. 
Formulations for the individual cross sections can be found in a number of 
sources~\citep{karzas_61,fernley_87,rumph_94,keady_00}. 

The total cross section, $\sigma$ is
\begin{eqnarray}
\sigma&=&N_{H^0}\sigma_H+N_{He^0}\sigma_{He^0}+N_{He^+}\sigma_{He}\\
	&=&\epsilon_H(1-x_H)\sigma_H+(1-\epsilon_{H})(1-x_{He^+}-x_{He^{+2}})\sigma_{He^0}
	+\epsilon_{He}x_{He^+}\sigma_{He}. \nonumber
	\label{eq:LymanXSection}
\end{eqnarray}

Here  $\epsilon_H$ and $\epsilon_{He}$  are the fractional abundances of hydrogen and
helium by number ($\epsilon_H+\epsilon_{He}\approx 1$), and $x_H$,  $x_{He^+}$ and $x_{He^{+2}}$ 
are the ionization fractions of H and He in the absorbing region.
We usually assume that all helium is neutral or singly ionized in the region of interest,
so that $x_{He^{+2}}=0$.

Possible ranges for these quantities would be:
 $0.1\ga x_H\ga 0.5$, $0.005\ga x_{He}\ga 0.14$, and $0.85\ga \epsilon_H\ga 0.95$~\citep{gilbert_05}. 

Continuum absorption in prominences was first noted in \textit{Skylab} data. 
Orrall \& Schmahl~\citep{orrall_76, orrall_80}, used absorption
measurements to determine column densities of neutral hydrogen. Similar
measurements were carried out with \textit{SOHO}~\citep{kucera_98, penn_00} and 
\textit{TRACE}~\citep{golub_99}, and have been performed using a number of different instruments
and combinations thereof. Typical H I column densities are on the order
$10^{18}-10^{19}$~cm$^{-2}$, but larger and smaller values have been reported.
Clearly, the exact values will depend on the particular feature analyzed and its
orientation. Gilbert et al.~\citep{gilbert_05,gilbert_06} first used absorption-based column density
measurements to estimate the mass for entire prominences. They found total
prominence mass values in the range $8\times10^{13}-2\times10^{15}$~g.

\begin{figure}
\includegraphics[width=8cm, trim=30 30 30 0]{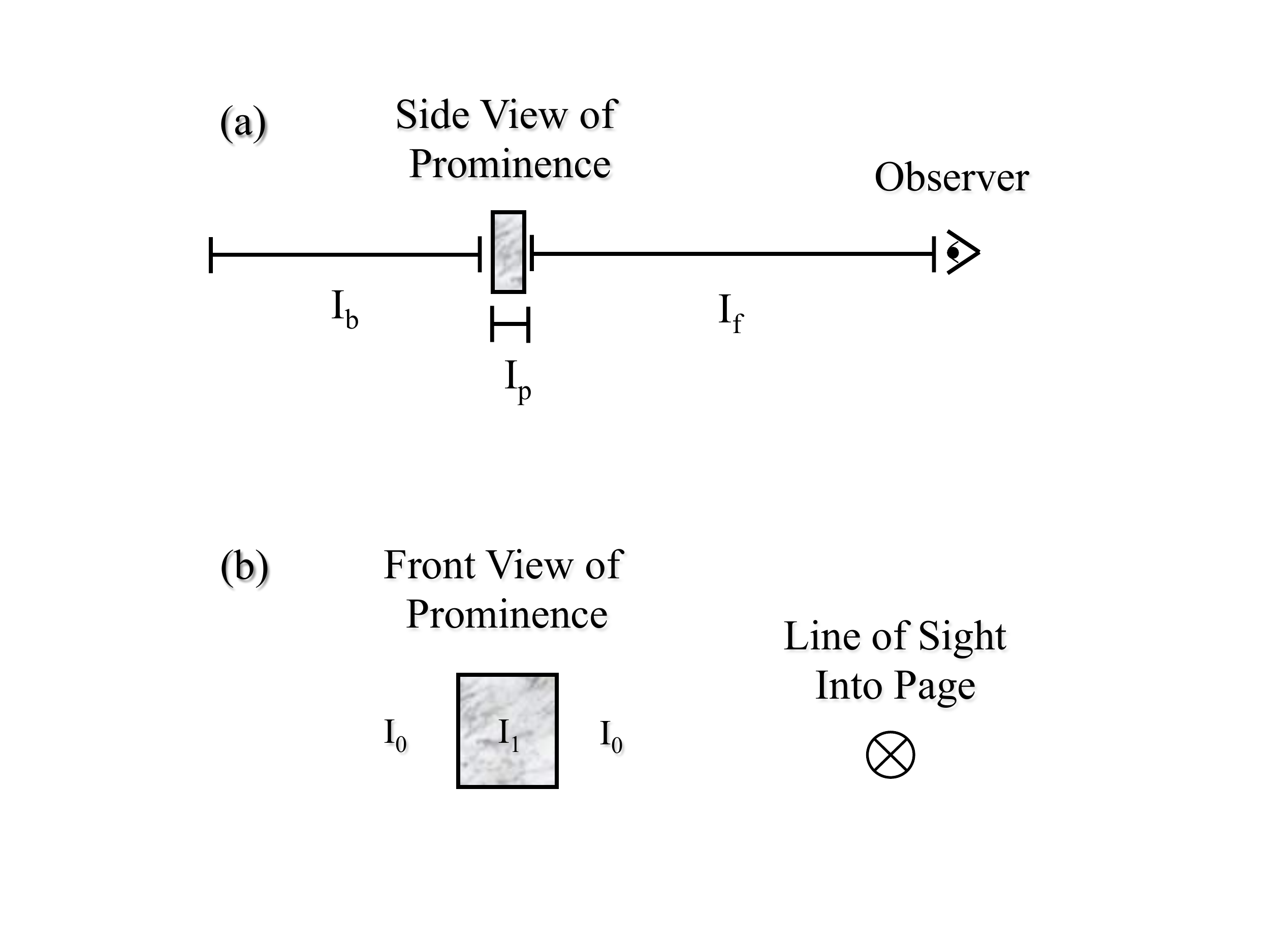}
\caption{Schematics showing the a prominence observed from (a) the side  and
(b) the front~\citep{gilbert_05}. (a) $I_b$ and
$I_f$ represent the emission in the background and foreground of the prominence
while $I_p$ is represents the coronal-temperature radiation from the region of the
prominence. (b) $I_0$ represents the observed intensity in areas of the sky without the prominence while
$I_1$ represents the intensity with the prominence.}
\label{f:LymanAbsGeom}
\end{figure}

The basic geometry of the problem is as shown in Fig.~\ref{f:LymanAbsGeom}. In
this formulation $I_0$ is the intensity that would be seen by the observer if
there were no prominence, while $I_1$ is the intensity in the portion of the sky 
containing the prominence material. In areas not including the prominence 
in the line of sight  these are equal. $I_0$ in the region 
containing the prominence is determined by either
interpolating spatially using adjacent areas without prominence material or
interpolating temporally if the prominence material is moving across the field
of view.

In each case the observed intensity is comprised of three parts along the line of sight: the intensity
in the foreground, $I_f$, the background, $I_b$ and from the region of the
prominence itself, $I_p$.
\begin{equation}
I_0=I_b+I_{0p}+I_f,
\label{eq:GHM1}
\end{equation}
\begin{equation}
I_1=\alpha I_b+I_{1p}+I_f.
\label{eq:GHM2}
\end{equation}
where the absorption is represented by the extinction factor, 

\begin{equation}
\alpha=\exp{\biggl(-\int^l_0 n\sigma ds\biggr)},
\label{eq:alpha}
\end{equation}
in which the product of the number density of particles and the absorption cross section 
in the source is integrated over the depth of the prominence along the line of sight, $l$.

The total mass is then 
\begin{equation}
M=(4(1-\epsilon_H)+\epsilon_H) m_H  \int\int -{\ln{\alpha}\over\sigma} da,
\end{equation}
where $m_H$ is the mass of hydrogen and $a$ is the area associated with the prominence absorption.

Thus there are at least five unknowns connected to the geometry alone. There are additional ones 
associated with the abundance and ionization parameters
(Eq.~\ref{f:LymanXSection})  and other issues like unresolved structure 
and portions of the prominence not detected in absorption.  
Researchers using continuum absorption techniques to calculate column
densities and prominence mass have approached these problems with a variety of
assumptions and combinations of data.

\textbf{Geometry}
For a prominence sufficiently in front of the plane of the sky (erupting from
near disk center, for instance) one might assume $I_f<<I_b$ and neglect the
foreground radiation all together. It has also been suggested that for a prominence 
on the limb one might assume $I_f\approx I_b$, although the corona is highly non-uniform, 
so such an assumption could be problematic.

Gilbert et al~\citep{gilbert_05} developed a technique for estimating $I_f$ and
$I_b$ by comparing the prominence emission against adjacent regions in which the
background but not the foreground emission is presumed to be varying (at the
solar limb, for instance).

\begin{figure}
\includegraphics[width=.8\textwidth]{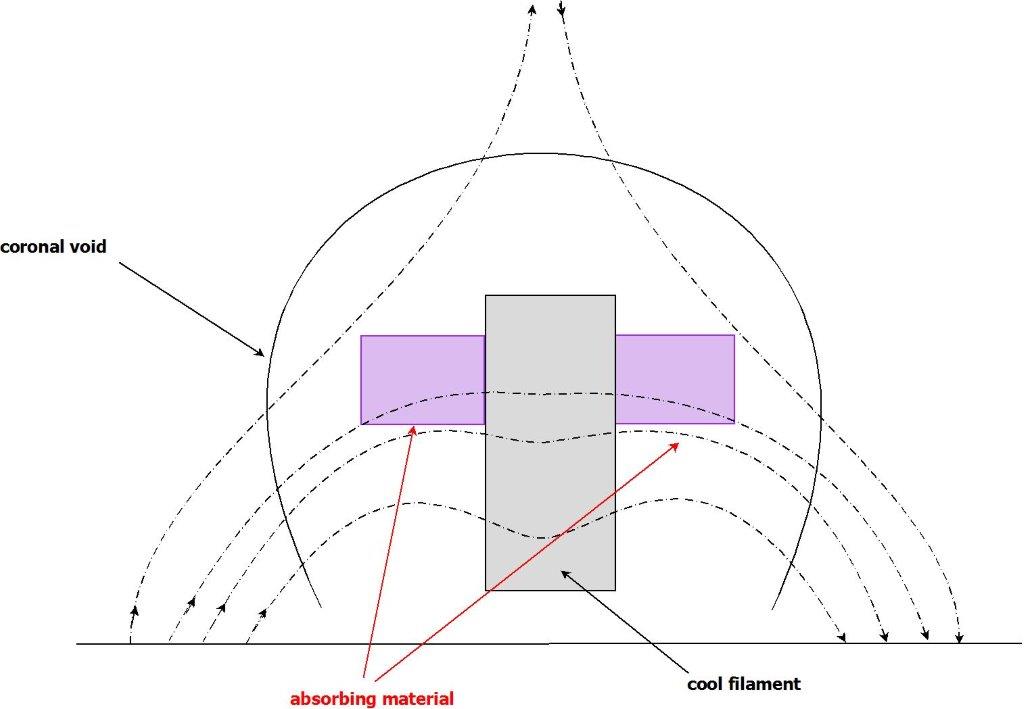}
\caption{Cartoon based on H-Ly$\alpha$ observations~\citep{vial_12} showing the location of the 
extended EUV filament  in a filament channel viewed along the inversion line. The central ``cool 
filament'' area shows the location of the prominence as would be seen in H$\alpha$. The ``absorbing 
material'' shows additional areas with material detected in Ly$\alpha$. Also shown is the outline of the 
coronal cavity (``coronal void'') and magnetic field lines (dashed lines).}
\label{f:ExtendedProm}
\end{figure}

\textbf{The Extended EUV Filament}
Some EUV observations reveal a more extended filament
structure than that often considered in the \Ha\ line. \Ha\ is similar in
optical depth to the continuum absorption near about 195~\AA~\citep{heinzel_08}. Thus,
as can be seen from Fig.~\ref{f:LymanXSection}, observations of filaments in
wavelengths that are significantly greater than 195~\AA\ but still less 
than the Lyman cut-off at 912  would
be expected to show more tenuous portions of the prominence. It is indeed the
case that EUV observations of filament channels, for instance in Mg~X 625~\AA,
show substantially wider filaments than those observed in \Ha~\citep{heinzel_03a,schwartz_04}.
Extended filament structure is also observed in hydrogen Lyman-$\alpha$ 
observations of prominences~\citep{vial_12,schwartz_12} and, as mentioned earlier in this chapter, prominences 
appear much larger in He II 304~\AA\ than in \Ha. 
Fig.~\ref{f:ExtendedProm} shows a suggested geometry for the extended filament based on H-Ly$\alpha$ observations.
It is not clear in such lines the extent to which we are observing more diffuse cool material or starting to observe the PCTR.
It has been estimated that the extra mass due to the filament extension may be equal to 50-100\% 
the mass derived from the \Ha\ emitting prominence~\citep{aulanier_02,heinzel_03a}. 

\textbf{Emissivity Blocking}
Emissivity Blocking (at one time referred to as volume blocking) is a reduction in the observed 
intensity in the corona due to the fact that cool structures do not emit radiation in lines formed 
at coronal temperatures~\citep{anzer_05, heinzel_08}.  This applies both to the situation
where the prominence is observable via absorption, and also to coronal line observations at wavelengths
not effected by continuum absorption or where such absorption is relatively weak. In these later 
cases there may be darkening due to emissivity blocking alone. 
Measurement of emissivity blocking can be performed with a combination of data
taken at wavelengths that do not exhibit continuum absorption and ones that do. These can be
combined to determine the amount of
darkening which occurs simply because of a lack of emission, allowing for a
accurate accounting for  the $I_{0p}$ and $I_{1p}$ terms in Eqs.~\ref{eq:GHM1} and \ref{eq:GHM2}.

\textbf{Substructure}
Unresolved sub-structure can also affect masses deduced from continuum
absorption~\citep{orrall_80, kucera_98,kucera_14}. One way to account for such sub-structure
is through a filling factor describing the fraction of the observing element that is filled 
with material. This quantity, $\fpos$, is an area filling factor and is distinct from the volume 
filling factor. Including it changes Eq.~\ref{eq:alpha} to:
\begin{equation}
\alpha=\fpos \exp{\biggl(-\int^l_0 n\sigma ds\biggr)}+ (1-\fpos)
\end{equation}
If $\fpos$ is low then the amount of material needed to account for a measured
amount of absorption, $\alpha$, will be higher than that needed if there is no
unresolved substructure.

\textbf{Prominence Emission}
It is also often assumed that $I_{1p}=0$, \ie\ that the emission from the prominence
itself is negligible. However, this may not be the case in some wave bands containing
contributions from the PCTR. In general when using continuum absorption 
it is best to select lines or wavebands in which there is no extra emission from
the prominence itself and substantial background emission.

\textbf{Other Uses for Continuum Absorption Measurements in Prominences}
Uncertainties in ionization fraction and helium abundance present difficulties for 
mass determinations using Lyman-continuum absorption. The converse of that  is that 
this absorption has potential for
measurements of abundances and temperature. Gilbert et al.~\citep{gilbert_11}
attempted to calculate the neutral H/He ratio by using \textit{SOHO} CDS data above
and below the 504 \AA\ for neutral helium absorption. The attempt was foiled by
optically thick absorption in one of the lines used (\ion{Mg}{10} at 625~\AA),
which is another thing to be careful of when analyzing absorption. However, the
measurement may be possible with a more sensitive instrument able to detect fainter
coronal lines and and fainter features. Landi \& Reale~\citep{landi_13} use the
dependence of the continuum absorption on ionization fraction and thus temperature to
estimate the temperature in an eruptive prominence.

\subsection{Cloud Modeling}
Mass can be determined with spectral lines by comparing observations with the
results of non-LTE magneto-hydrostatic models~\citep{mein_96,heinzel_99,heinzel_15}. 
By measuring or making assumptions about temperature,
bulk velocity and turbulent velocity it is possible use a grid of model results
combined with measurements of the filament dimensions to calculate the filament
mass. As with the continuum absorption method, uncertainties in ionization fractions
result in significant uncertainties in the mass values. Masses calculated in
this way give values in the range $2 - 6 \times10^{15}$~g, on the order of values obtained using
calculations based on continuum absorption~\citep{koutchmy_08,grechnev_14}.

\subsection{White Light Measurements}
Erupting prominences contribute to the mass calculated  for CMEs
done using white light coronagraph observations. If the emission observed is purely
due to Thompson scattering off electrons  then mass is proportional to the
electron density with an angle dependence emphasizing the contribution in the
plane of the sky. Inversions of the white light intensity to
calculate the CME mass have been performed since the 1970s~\citep{stewart_74}.

To calculate CME mass the white light associated with the CME is determined by subtracting off a pre-event image. 
The mass is then~\citep{vourlidas_10b}:
\begin{equation}
M_{\textit{\footnotesize{CME}}}=I_{\textit{\footnotesize{CME}}}\, C_e\, C_{\textit{\footnotesize {plasma}}},
\end{equation}
where $I_{\textit{\footnotesize{CME}}}$ is the excess brightness associated with the CME in units of mean 
solar brightness and $C_{\textit{\footnotesize{plasma}}}$ is the composition of the CME plasma. If one 
assumes a composition of 10\% helium   $C_{\textit{\footnotesize{plasma}}}$ is $2\times10^{-24}$~g~electron
$^{-1}$.  $C_e$ is obtained from Thompson scattering theory~\citep{billings_66chp6B,hundhausen_93} using 
the assumption that all electrons lie in the plane of the sky. The mass is determined for each pixel and 
summed over the area associated with the CME.

It is also usually assumed that the prominence material still present is completely ionized ~\citep{vourlidas_10b}. 
If that is not the case it can result in an over estimate of CME mass due to \Ha\ emission in the white 
light bandpass. Athay and Illing~\citep{athay_86} took into account both white light and \Ha\ to calculate 
the mass of an erupting prominence, including an analysis of the ionization state of hydrogen, to 
obtain a mass of $1\times10^{16}$~g. 
 
 \section{Some Outstanding Questions Related to Prominence Flows and Mass}
 There are numerous questions concerning prominences to which measurements of flows and 
 mass are key. Most of them tie into the basic question of the origins of prominence mass. Clearly we 
 need to bring all available observations of flows and oscillations along with
 measurements of temperature, mass, density, and magnetic field  and, of course, modeling 
 to bear on this larger question and the related questions listed below.
  
 \begin{itemize}
\item{Where and at what temperature does prominence plasma originate? This is an
important distinguisher between different theories of the origin of prominence
mass. Some observations, especially in active regions, suggest that prominence material
originates as relatively cool plasma ejected from the chromosphere~\citep{wangym_99,schmieder_04}. Other
observations suggest that the cool material condenses from hot material~\citep{berger_12}. These 
differences may be the result of more than one mechanism at work in different situations, 
for instance between active region vs.\ quiescent prominences, but this still is not entirely clear.}
\item{In a related question, what is the role of barbs? Are they the chief conduit of material
to and from the chromosphere, important to the existence of prominence plasma, or
are they only the result of perturbations on the prominence channel
magnetic field with no special role in the processes moving material into the
prominence?}
\item{The complex flows in hedgerow prominences offer a fascinating 
challenge, offering us a look into a regime in which apparently different 
processes dominate than in lower latitude prominences.  Further detailed observations of these motions 
in three dimensions (spectra with high spatial and temporal resolution 
combined with high quality imaging) and magnetic field information~\citep{orozco_14} should help us understand the processes at work.}
\item{How are the plasmas in the prominence and the larger prominence
cavity related?  Three dimensional morphological modeling using EUV images and 
coronal magnetic field measurements could help us understand the prominence cavity system.}
\item{What determines the size and structure of prominence flows at different scales? 
Substructure may reflect the manner in which the material is inserted into the corona~\citep{kucera_14}, or, 
alternatively, be a result of processes occurring in the prominence itself, including evolution of the 
magnetic field or wave based variations~\citep{antolin_14}.}

 \end{itemize}
 
The understanding of flows and mass distributions in prominences is central to these and other 
important questions about the nature of these complicated and mysterious features of the solar atmosphere.

\bibliographystyle{plain}

\end{document}